\documentclass[8.5pt,twocolumn]{article}
\usepackage[super,sort&compress,comma]{natbib}
\usepackage{mhchem}
\usepackage{times,mathptmx}
\usepackage{secsty}
\usepackage{balance}

\usepackage{graphicx} %eps figures can be used instead
\usepackage{lastpage}
\usepackage[format=plain,justification=raggedright,singlelinecheck=false,font=small,labelfont=bf,labelsep=space]{caption}
\usepackage{fancyhdr}
\usepackage{color}
\usepackage{amsmath} % AMS Math Package
\usepackage{amsthm} % Theorem Formatting
\usepackage{amssymb}	% Math symbols such as \mathbb
\usepackage{graphicx} % Allows for eps images
\usepackage{multicol} % Allows for multiple columns
\usepackage[dvips,letterpaper,margin=0.75in,bottom=0.5in]{geometry}
\makeatletter % Need for anything that contains an @ command
\renewcommand{\maketitle} % Redefine maketitle to conserve space
{ \begingroup \vskip 10pt \begin{center} \large {\bf \@title}
	\vskip 10pt \large \@author \hskip 20pt \@date \end{center}
  \vskip 10pt \endgroup \setcounter{footnote}{0} }
\makeatother % End of region containing @ commands
\newcommand{\ket}[1]{\left| #1 \right>} % for Dirac bras
\let\baraccent=\= % rename builtin command \= to \baraccent
\renewcommand{\=}[1]{\stackrel{#1}{=}} % for putting numbers above =

\theoremstyle{definition}

\theoremstyle{remark}

\begin{document}

\pagestyle{fancy}
\thispagestyle{plain}
\fancypagestyle{plain}{
%\fancyhead[L]{}
%\fancyhead[C]{}
%\fancyhead[R]{}
\renewcommand{\headrulewidth}{1pt}}
\renewcommand{\thefootnote}{\fnsymbol{footnote}}
\renewcommand\footnoterule{\vspace*{1pt}%
\hrule width 3.4in height 0.4pt \vspace*{5pt}} \setcounter{secnumdepth}{5}

\makeatletter
\def\subsubsection{\@startsection{subsubsection}{3}{10pt}{-1.25ex plus -1ex
minus -.1ex}{0ex plus 0ex}{\normalsize\bf}}
\def\paragraph{\@startsection{paragraph}{4}{10pt}{-1.25ex plus -1ex minus
-.1ex}{0ex plus 0ex}{\normalsize\textit}}
\renewcommand\@biblabel[1]{#1}
\renewcommand\@makefntext[1]%
{\noindent\makebox[0pt][r]{\@thefnmark\,}#1} \makeatother
\renewcommand{\figurename}{\small{Fig.}~}
\sectionfont{\large}
\subsectionfont{\normalsize}

\fancyfoot{}
%\fancyfoot[LO,RE]{\vspace{-7pt}\includegraphics[height=9pt]{headers/LF}}
%\fancyfoot[CO]{\vspace{-7.2pt}\hspace{12.2cm}\includegraphics{headers/RF}}
%\fancyfoot[CE]{\vspace{-7.5pt}\hspace{-13.5cm}\includegraphics{headers/RF}}
\fancyfoot[RO]{\footnotesize{\sffamily{1--\pageref{LastPage} ~\textbar
\hspace{2pt}\thepage}}}
\fancyfoot[LE]{\footnotesize{\sffamily{\thepage~\textbar\hspace{3.45cm}
1--\pageref{LastPage}}}} 
\fancyhead{}
\renewcommand{\headrulewidth}{1pt}
\renewcommand{\footrulewidth}{1pt}
\setlength{\arrayrulewidth}{1pt} \setlength{\columnsep}{6.5mm}
\setlength\bibsep{1pt}

\twocolumn[
  \begin{@twocolumnfalse}
	\noindent\LARGE{\textbf{Molecular spectroscopy for ground-state transfer of
ultracold RbCs molecules}} \vspace{0.6cm}

\noindent\large{Markus Debatin\textit{$^{a}$}, Tetsu
Takekoshi\textit{$^{a,b}$}, Raffael Rameshan\textit{$^{a}$}, Lukas
Reichs\"{o}llner\textit{$^{a}$}, Francesca Ferlaino\textit{$^{\ast a}$}, Rudolf
Grimm\textit{$^{a,b}$}, Romain Vexiau\textit{$^c$}, Nadia Bouloufa\textit{$^c$}, Olivier Dulieu\textit{$^c$}, and Hanns-Christoph N\"{a}gerl\textit{$^{a}$}}\vspace{0.5cm}

%Please note that \ast indicates the corresponding author(s) but no footnote %text is required.

\noindent \normalsize{We perform one- and two-photon high resolution
spectroscopy on ultracold samples of RbCs Feshbach molecules with the aim to
identify a suitable route for efficient ground-state transfer in the
quantum-gas regime to produce quantum gases of dipolar RbCs ground-state
molecules. One-photon loss spectroscopy allows us to probe deeply bound
rovibrational levels of the mixed excited $(A^1\Sigma^+ - b^3\Pi_{0}) \ 0^+$
molecular states. Two-photon dark state spectroscopy connects the initial
Feshbach state to the rovibronic ground state. We determine the binding energy of the lowest rovibrational level $\ket{v''\!=\!0,J''\!=\!0}$ of the $X^1\Sigma^+$ ground state to be $D_0^X=3811.5755(16)$~cm$^{-1}$, a 300-fold improvement in accuracy with respect to previous data. We are now in the position to perform stimulated two-photon Raman transfer to the rovibronic ground state.}
\vspace{1.5cm}
\end{@twocolumnfalse}
] %Footnotes %\footnotetext{\dag~Electronic Supplementary Information (ESI) available:
%[details of any supplementary information available should be included
%here]. See DOI: 10.1039/b000000x/}

%Please use \dag to cite the ESI in the main text of the article.
%If you article does not have ESI please remove the the \dag symbol from the %title and the above footnotetext.

\footnotetext{\textit{$^{a}$~ Institut f\"ur Experimentalphysik und Zentrum
f\"ur Quantenphysik, Universit\"at Innsbruck, Technikerstrasse 25, A-6020
Innsbruck, Austria.\\E-mail: Francesca.Ferlaino@uibk.ac.at}}
\footnotetext{\textit{$^{b}$~Institut f\"ur Quantenoptik und
Quanteninformation, \"Osterreichische Akademie der Wissenschaften, A-6020
Innsbruck, Austria}} \footnotetext{\textit{$^{c}$~Laboratoire Aim\'e Cotton,
CNRS, Universit\'e Paris-Sud, B\^at. 505, 91405 Orsay Cedex, France.\\E-mail:
olivier.dulieu@u-psud.fr}}

%additional addresses can be cited as above using the lower-case letters, c, d, e...
%If all authors are from the same address, no letter is required

%\footnotetext{\ddag~Additional footnotes to the title and authors can be
%included \emph{e.g.}\ `Present address:' or `These authors contributed %
%equally to this work' as above using the symbols: \ddag, \textsection,
%and \P. Please place the appropriate symbol next to the author's name
%and include a \texttt{\textbackslash footnotetext} entry in the the correct
%place in the list.}

\section*{Introduction}
\label{sec:intro} Ultracold and dense molecular samples promise myriad
possibilities for new directions of research in fields such as ultracold
chemistry, quantum information science, precision metrology, Bose-Einstein
condensation, and quantum many-body physics. Recent overviews on the status of
the field and on possible experiments can be found in Refs.
\cite{Carr2009cau,Friedrich2009wac,dulieu2009} Within our work on ultracold
molecules we are primarily interested in dipolar many-body physics.
Heteronuclear molecules such as KRb and RbCs in their rovibronic ground state
feature sizable permanent electric dipole moments of about
1~Debye,\cite{Kotochigova2005air,Deiglmayr2008cos} giving rise to long range
and anisotropic dipole-dipole interactions in the presence of external
polarizing electric fields. As a consequence, at ultralow temperatures,
few-body and many-body dynamics are expected to be governed by dipolar effects. In particular, a broad variety of novel quantum phases has been proposed for
bosonic and fermionic dipolar gases that are confined to two- or three-dimensional lattice potentials.\cite{Goral2002qpo,Barnett2006qmw,Menotti2007mso,Buchler2007sc2,
Yi2007nqp,Danshita2009sos,Cooper2009sts,Capogrosso2010qpo,Pollet2010spw,
Potter2010sad,Pikovski2010isi,Li2010cmo}

Our goal is to produce a quantum gas of ground-state RbCs molecules. The molecule RbCs, in its lowest internal state, is stable under binary collisions,\cite{Zuchowski2010} i.e.~the exchange reaction 2RbCs $\rightarrow$ Rb$_2$ + Cs$_2$ is not possible (this is e.g. not the case for the molecule KRb). Collisional stability is an important requirement for the formation of a molecular quantum gas in full thermal equilibrium. As molecular samples are not readily laser cooled, we intend to use a scheme in which the molecules are produced at high phase-space density out of an atomic Rb-Cs quantum-gas mixture\cite{Lercher2011poa} by Feshbach association and coherent state-transfer techniques, similar to recent work on Cs$_2$,\cite{Danzl2008qgo,Danzl2010auh} Rb$_2$,\cite{Lang2008utm} and
KRb.\cite{Ni2008ahp} Experiments on KRb have reached a regime in which dipolar effects on reaction dynamics and thermalization processes have become very pronounced.\cite{Ni2010dco,Ospelkaus2010qcc,deMiranda2011} For our experiments on RbCs an optical lattice will aid in the association and state-transfer process by preventing the molecules from undergoing collisional loss before they have reached the rovibrational ground state. Initially, RbCs Feshbachmolecules\cite{Pilch2009ooi,Takekoshi2011} will be formed out of two-atom Mott-insulator state in a generalized version of previous work on the efficient formation of homonuclear Rb$_2$ and Cs$_2$ molecules.\cite{Volz2006pqs,Danzl2010auh} The molecules will then be transferred to the lowest rovibronic level of the $X^1\Sigma^+$ absolute ground state on a two-photon transition using the STImulated Raman Adiabatic Passage
(STIRAP) technique.\cite{Bergmann1998cpt}

Here, we perform high-precision molecular spectroscopy on a dense sample of
ultracold RbCs molecules to identify a suitable two-photon path to the
rovibronic $X^1\Sigma^+$ ground state as depicted in Fig.~\ref{fig:stirap_scheme} (a). Starting from a Feshbach level $\ket{1}$, we
perform molecular loss spectroscopy to search for and identify a suitable level $\ket{2}$ of the mixed $(A^1\Sigma^+ - b^3\Pi_{0}) \ 0^+$ excited molecular
potentials, hereafter named $0^+$ potentials. We probe levels that lie 6320 to 6420~cm$^{-1}$ (1580~nm to 1557~nm) above the Rb($5^2S$)+Cs($6^2S$) limit and hence 10160 to 10260~cm$^{-1}$ (980~nm to 970~nm) above the rovibronic ground-state level $\ket{3}$. The wavelength ranges above were chosen because they feature sufficiently strong transitions and because they can be covered by readily available diode laser sources. We then perform dark-state spectroscopy by simultaneous laser irradiation near 1570~nm and 980~nm, connecting the initial Feshbach level $\ket{1}$ to the rovibronic ground state $\ket{3}$. We find several dark resonances, from which we derive normalized transition strengths and find that some of the two-photon transitions are candidates for
ground-state transfer. The paper is organized as follows: In Section
\ref{sec:molpot} we describe the molecular structure relevant to identify a
suitable route for the production of ultracold samples of RbCs molecules in
their rovibronic ground state. Section \ref{sec:FeshMol} summarizes the sample
preparation procedure. Section \ref{sec:0plus} then presents the results of our one-photon loss spectroscopy measurements, while Section \ref{sec:TwoPhot}
discusses the results of the two-photon dark state spectroscopy. We finally
give an outlook on future work.

\section{The route towards RbCs ground-state molecules}
\label{sec:molpot}

Our strategy to populate the rovibronic ground state is based on a STIRAP
transfer scheme.\cite{Bergmann1998cpt} This method is coherent, highly
efficient, robust, fast, reversible, and allows control over all the
internuclear degrees of freedom including vibration, mechanical rotation, and
nuclear spin. STIRAP typically involves three molecular levels that are coupled by a two-photon transition in a lambda-type configuration. To achieve highly
efficient STIRAP transfer it is mandatory to determine a suitable three-level
system. The initial state $\ket{1}$ at threshold is a weakly bound Feshbach
state. This state happens to be of almost pure triplet $a^3\Sigma^+$ character. We populate this state out of an ultracold Rb-Cs mixture via the Feshbach
association technique\cite{Herbig2003poa} as discussed below. The intermediate
level $\ket{2}$ is a level belonging to the electronically excited
$0^+$ potentials mixed by spin-orbit interaction. The final state $\ket{3}$ is the lowest rovibrational level of the $X^1\Sigma^+$ ground state potential. Laser $L_1$ connects the initial Feshbach level $\ket{1}$ to the intermediate level $\ket{2}$, and laser $L_2$ couples the latter to $\ket{3}$.

Our goal is to identify a suitable intermediate level $\ket{2}$ from the
$0^+$ potentials with sufficient oscillator strength with both levels $\ket{1}$ and $\ket{3}$. The STIRAP process takes advantage of the mixed singlet-triplet character of the $0^+$ pair of states correlated to the Rb-Cs dissociation limit $5^2S_{1/2}+6^2P_{1/2,3/2}$ (see also Table~\ref{tab:spectro} below). Here $\Omega'=0$ labels the projection of the total electronic angular momentum on the molecular axis, and the plus sign the parity of the state for the symmetry with respect to a plane containing the molecular axis. These two states result from the spin-orbit interaction between the lowest state of $^3\Pi$ symmetry and the second state of $^1\Sigma^+$ symmetry in RbCs correlated to the Rb($5^2S$)+Cs($6^2P$) limit, hereafter referred to as the $b^3\Pi$ and $A^1\Sigma^+$ states. Previous spectroscopic analysis\cite{Docenko2010} of these states has resulted in two potential curves for the $b^3\Pi$ and $A^1\Sigma^+$ states and molecular spin-orbit coupling functions varying with the internuclear distance $R$. A potential curve for the $\Omega'\!=\!1$ component of the fine structure manifold of the $b^3\Pi$ state is also provided by Ref.\cite{Docenko2010} We chose the $X^1\Sigma^+$ ground state potential curve determined in Ref.\cite{Fellows1999} by the Inverted Perturbation Approach (IPA) of spectroscopic data. In principle, level $\ket{1}$ is determined by the details of the hyperfine structure of the ground-state manifold. With its nearly pure triplet character it can be described as if it belonged to a single $a^3\Sigma^+$ potential curve for the purpose of our model. No experimental determination of this potential
is available yet, although we are aware of ongoing work in the group of D.
DeMille at Yale University. We use the \textit{ab initio} potential calculated
in Ref.\cite{Aymar2005}, matched at 22~a.u. to the asymptotic potential derived from the spectroscopic analysis of the $X^1\Sigma^+$ ground state.\cite{Fellows1999} We modify the $C_6$ van der Waals coefficient to the
value $C_6=5663$~a.u.\cite{Derevianko2001} In addition we slightly moved the
repulsive wall to reproduce the scattering length of 542~a.u. reported in
Ref.\cite{Takekoshi2011} This corresponds to a shift of the inner turning point at zero energy of about 0.2~a.u.

\begin{figure}[h]
\begin{center}
\includegraphics[width=0.45\textwidth]{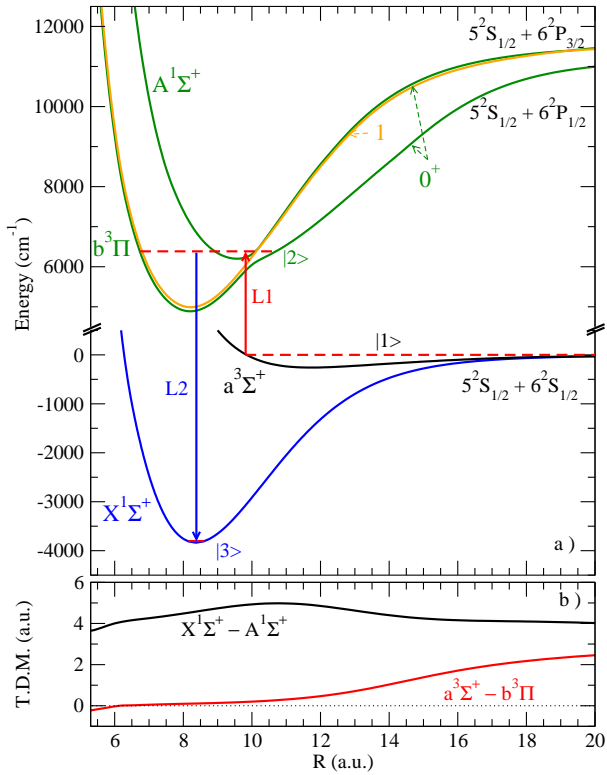}
\caption{\label{fig:stirap_scheme} \small  Proposed STIRAP scheme to transfer
RbCs Feshbach molecules to the lowest rovibrational level of the ground state.
(a) Level $ \ket{1} $ is the initial Feshbach level with mostly $a^3\Sigma^+$
character. Level $\ket{3}$ is the $\ket{v"\!=\!0, J"\!=\!0}$ level of the
$X^1\Sigma^+$ ground state. Level $\ket{2}$ belongs to the $b^3\Pi(0^+)$ and
$A^1\Sigma^+(0^+)$ coupled states. The wavelengths of lasers L$_1$ and L$_2$
are taken here as 1569~nm and 982~nm. (b) Transition dipole moments (TDM)
$\mu^{a-\Pi}(R)$ and $\mu^{X-\Sigma}(R)$ as functions of the internuclear
distance $R$. The potential curve for the $\Omega'=1$ component of the fine
structure manifold of the $b^3\Pi$ state is also displayed.}
\end{center}
\end{figure}

\begin{figure}[h]
\begin{center}
\includegraphics[width=0.45\textwidth]{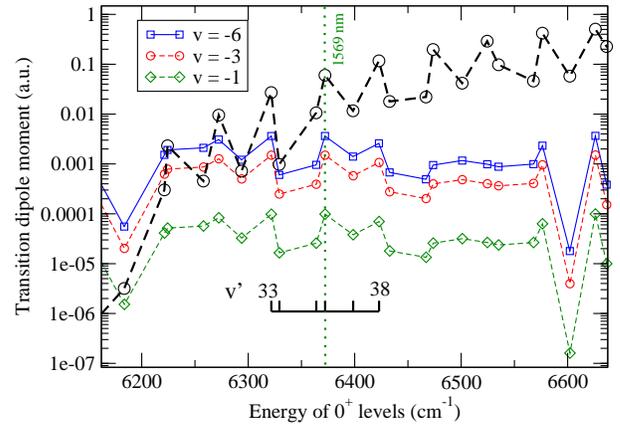}
\caption{\label{fig:dip_vibration} \small Transition dipole moments without rotation $\mu^{X-0^+}_{v''v'}$ (dashed black line with open circles) and $\mu^{a-0^+}_{vv'}$ for various levels $v=-1, -3, -6$ of the $a^3\Sigma^+$ potential numbered downward from the dissociation limit. The energy is taken from the Rb($5^2S_{1/2}$)+Cs($6^2S_{1/2}$) limit without hyperfine structure. The vertical line shows the laser wavelength relevant for the experiment. A ratio of 2.42$\pm 0.02$ and 37$\pm 0.5$ is found between the TDMs for $v=-6$ and $v=-3$, and for $v=-6$ and $v=-1$, respectively.}
\end{center}
\end{figure}

Using the Mapped Fourier Grid Hamiltonian (MFGH) method\cite{Kokoouline1999} we compute the vibrational wave functions $\Psi^{(X)}(R)$, $\Psi^{(a)}(R)$, and
$\Psi^{(0^+)}(R)$ of the $X^1\Sigma^+$, $a^3\Sigma^+$, and $0^+$ molecular
states, respectively, ignoring rotation in a first step. The upward STIRAP step relies on the $a^3\Sigma^+ \rightarrow b^3\Pi$ electronic transition, and the downward step relies on the $A^1\Sigma^+ \rightarrow X^1\Sigma^+$ transition. The relevant $R$-dependent transition dipole moment functions are calculated following Ref.\cite{Aymar2005} We plot them in Fig.\,\ref{fig:stirap_scheme}(b). The resulting transition dipole moments (TDMs) $\mu^{a-0^+}_{vv'}=\left \langle \Psi^a_v  | \mu^{a-\Pi}(R)  |
\Psi^{0^+(\Pi)}_{v'} \right \rangle$ and $\mu^{X-0^+}_{v''v'}= \left \langle
\Psi^X_{v''}  | \mu^{X-\Sigma}(R)  | \Psi^{0^+(\Sigma)}_{v'} \right \rangle$
are displayed in Fig.\,\ref{fig:dip_vibration} as functions of the $0^+$ level
energies. Here, $v$, $v'$ and $v''$ are the vibrational quantum numbers of the initial, intermediate, and final levels, respectively. Note that $v'$ labels  the levels of the $0^+$ coupled states corresponding to their numbering with increasing energy, which is not regular in contrast to the case of a single regular potential well. First, we find that the $\mu^{a-0^+}_{vv'}$ TDMs exhibit identical variations for $v$ levels close
to the dissociation limit apart from a $v$-dependent global scaling factor
related to the amplitude of the wave function at the inner turning point of the $a^3\Sigma^+$ potential. This simply expresses the fact that the TDMs are
mainly influenced by the inner part of the initial wave function where
absorption occurs rather than by its multichannel character at large distances, and thus justifies the present model. Second, $\mu^{X-0^+}_{v''v'}$ is much
larger than $\mu^{a-0^+}_{vv'}$ above 6400~cm$^{-1}$, reflecting the almost
vanishing transition dipole moment around the $a^3\Sigma^+$ inner turning
point. Below this energy the spatial overlaps of $v'$ and $v''=0$ wave functions rapidly decrease as the bottoms of the $X$ and $A$ wells are shifted from each
other.

In order to understand the STIRAP selection rules, levels $\ket{1}$, $\ket{2}$, and $\ket{3}$ must be expressed in terms of a common quantum number $J$, where
$\vec{J}=\vec{L}+\vec{S}+\vec{\ell}$ is the total angular momentum of the
molecule ignoring nuclear spin. Here, $\vec{L}+\vec{S}$ is the total electronic angular momentum and $\vec{\ell}$ is the pure rotation of the molecule. In the
case of levels $\ket{2}$ and $\ket{3}$ the number $J$ determines the rotational energy level structure of the molecule, while it is not well defined for the Feshbach level $\ket{1}$. For details, see Section \ref{sec:0plus}. Assuming that the radial wave function of a rovibrational level $(v,J)$ is independent of $J$ for the low $J$ values involved, the transition dipole moments including rotation $D^{q,\pm \epsilon}_{\ket{i},\ket{f}}$ between the initial $(v_i,J_i,M_i)$ and the final level $(v_f,J_f,M_f)$ of the electronic states $\Gamma_i$ and $\Gamma_f$, respectively, are related to the purely vibrational dipole moments $\mu^{\Gamma_i-\Gamma_f}_{v_i,v_f}$ using the usual H\"onl-London factors
\begin{eqnarray*}
D^{q,\pm \epsilon}_{\ket{i},\ket{f}} &=&
\sqrt{\frac{(2J_i+1)(2J_f+1)(1+q)}{1+|\epsilon|}}\\ \nonumber
&\times&\left(\begin{array}{ c c c }
    J_f & 1 & J_i \\
    -M_{f} & \epsilon & M_{i} \\
  \end{array} \right)\\ \nonumber
&\times& \left(\begin{array}{ c c c }
    J_f & 1 & J_i \\
    -\Lambda_f  & q & \Lambda_i\\
  \end{array} \right) \times \mu^{\Gamma_i-\Gamma_f}_{v_i,v_f},
\end{eqnarray*}

where $M_i$ ($M_f$) is the projection of $J_i$ ($J_f$) onto the lab-frame
quantization axis and $\Lambda_i$ ($\Lambda_f$) is the projection onto the
molecular axis. The index $q$ labels $\Sigma-\Sigma$ ($q=0$) and $\Sigma-\Pi$
($q=\pm 1$) transitions, and $\epsilon=0$ ($=\pm 1$) defines the laser
polarization along (perpendicular to) the quantization axis. For the purpose of the next sections we present in Table \ref{tab:Honl-London} the factors related to the possible values of the projection of the angular momentum of level $
\ket{1} $, i.e. $M_J=0,\pm 1, \pm 2$ for the two accessible rotational states
$J'=1,3$ of the $ \ket{2} $ levels.

\begin{table}[h]
\begin{center}
 \begin{tabular}{|c|c|c|c|c|c|c|} \hline
 \small
& \multicolumn{3}{c|}{J'=1} & \multicolumn{3}{c|}{J'=3} \\ \cline{2-7}
   $M_J$ & v & \multicolumn{2}{c|}{h} & v & \multicolumn{2}{c|}{h} \\
  & & $\sigma_-$ &$\sigma_+$ & &$\sigma_-$ & $\sigma_+$ \\   \cline{3-6}
  \hline
  &&&&&&\\
$0$
&$\sqrt{\frac{2}{15}}$&$\sqrt{\frac{1}{60}}$&$\sqrt{\frac{1}{60}}$&$\sqrt{\frac{12}{35}}
$&$\sqrt{\frac{4}{35}}$&$\sqrt{\frac{4}{35}}$ \\ &&&&&&\\
$+1$&$\sqrt{\frac{1}{10}}$&$\sqrt{\frac{1}{20}}$&        0
&$\sqrt{\frac{32}{105}}$&$\sqrt{\frac{2}{35}}$&$\sqrt{\frac{4}{21}}$ \\&&&&&&
\\ $-1$&$\sqrt{\frac{1}{10}}$&        0
&$\sqrt{\frac{1}{20}}$&$\sqrt{\frac{32}{105}}$
&$\sqrt{\frac{4}{21}}$&$\sqrt{\frac{2}{35}}$ \\&&&&&& \\ $+2$&        0
&$\sqrt{\frac{1}{10}}$&        0
&$\sqrt{\frac{4}{21}}$&$\sqrt{\frac{2}{105}}$&$\sqrt{\frac{2}{7}} $ \\&&&&&& \\
$-2$&        0            &        0
&$\sqrt{\frac{1}{10}}$&$\sqrt{\frac{4}{21}}$&$\sqrt{\frac{2}{7}}
$&$\sqrt{\frac{2}{105}}$ \\&&&&&& \\ \hline
\end{tabular}
\caption{\label{tab:Honl-London} \small H\"onl-London factors for the
transitions between the rotational state $J=2, M_J$ of level $ \ket{1} $ and
the rotational states $J'=1$ and $J'=3$ of level $ \ket{2} $. The labels "v"
and "h" refer to the polarization $\epsilon=0$ and $\pm 1$, respectively.}
\end{center}
\end{table}

\section{Experimental setup and initial state preparation}
\label{sec:FeshMol}

The starting level $\ket{1}$ for optical spectroscopy and future STIRAP
experiments is a Feshbach molecular level that we created directly from an
$^{87}$Rb-$^{133}$Cs atomic mixture by using the standard magneto-association
technique using a Feshbach
resonance.\cite{Herbig2003poa,Xu2003foq,Durr2004oom,Ferlaino2009ufm,Chin2010fri}
We produce a pure sample of RbCs Feshbach molecules trapped in an optical
dipole trap as follows. We first prepare an atomic mixture following an
all-optical scheme similar to the one described in Ref.\cite{Pilch2009ooi,
Lercher2011poa} In brief, Rb and Cs atoms are captured in a two-species
magneto-optical trap (MOT) from a two-species Zeeman slower. After a stage of
two-color degenerate Raman-sideband cooling\cite{Pilch2009ooi}, a technique
that simultaneously cools and polarizes the mixture in the absolute lowest
hyperfine Zeeman level, the atoms are loaded into a magnetically levitated
large-volume dipole trap. After 1~s of plain evaporation the dipole trap
typically contains 1.7 $\times 10^6$ Rb and Cs atoms at about 5$\mu$K in the
$\ket{f_\text{Rb}\!=\!1, m_{f_\text{Rb}}\!=\!1\!}$ and  $|f_\text{Cs}\!=\!3,
m_{f_\text{Cs}}\!=\!3\!\rangle$ Zeeman sublevels, respectively. As usual
$f_\text{Rb}$ ($f_\text{Cs}$) is the total atomic angular momentum quantum
number and $m_{f_\text{Rb}}$ ($m_{f_\text{Cs}}$) its projection onto the
magnetic field axis. While a large volume trap is well suited for optimal
loading of atoms from the MOT, a tightly focused dipole trap is desirable to
increase the number density and to assure the high thermalization rate required for evaporative cooling towards Bose-Einstein condensation. However, Rb and Cs
have a comparatively large interspecies background scattering length as a
result of the presence of a near threshold $s$-wave state belonging to the open scattering channel.\cite{Takekoshi2011} The large scattering length is
responsible for the large three-body Rb-Cs inelastic collisional rate observed
in our recent experiments.\cite{Lercher2011poa} We follow the strategy of
spatially separating the two species by loading Rb and Cs atoms into two
different tightly focused dipole traps.\cite{Lercher2011poa} We then perform
evaporative cooling on each species separately. At the onset of condensation,
we stop the forced evaporation and we superimpose the Cs atomic cloud onto the
Rb one by spatially shifting the Cs dipole trap. We end up with an atomic mixture of $1.5\times 10^5$ Rb and $0.8\times 10^5$ Cs atoms at a temperature of about 200~nK.

The near-degenerate atomic mixture is the starting point for Feshbach
association.\cite{Takekoshi2011} In previous experiments we have found a large
number of Feshbach resonances in the magnetic field range up to about
670~G.\cite{Pilch2009ooi,Takekoshi2011} In principle, each observed resonance
can be used to convert atoms into molecules. For our one-photon spectroscopic
measurements we have to match both the requirement of high conversion
efficiency and large wave-function overlap between the starting molecular
Feshbach level and the final molecular state in the electronically excited
mixed $0^+$ potentials. We identify two different Feshbach levels to be the most appropriate candidates for the initial state $\ket{1}$. The two levels intersect the Rb($5^2S_{1/2},f_\text{Rb}\!=\!1$)+Cs($6^2S_{1/2},f_\text{Cs}\!=\!3$) atomic
threshold at a magnetic field strength $B$ of about $217$\,G and  $182$\,G,
referred to in the following as $\ket{1\rm{a}}$ and $\ket{1\rm{b}}$,
respectively. We populate these levels by the usual magnetic field ramping
technique.\cite{Chin2010fri,Takekoshi2011} To avoid undesired atom-molecule
collisions we remove the Rb and Cs atoms by Stern-Gerlach
separation.\cite{Takekoshi2011} We finally obtain an optically trapped sample
of up to 4000 RbCs Feshbach molecules at a temperature of about 200~nK. To
image the molecules and to determine their number we ramp back across the
Feshbach resonance to dissociate the molecules and take separate absorption
pictures of the Rb and Cs atomic samples.\cite{Lercher2011poa}

The Feshbach molecular states near threshold are typically described using the
atomic basis. The full set of good quantum numbers is given by
$\ket{\tilde{v}; f_\text{Rb}, m_{f_\text{Rb}}; f_\text{Cs}, m_{f_\text{Cs}};
m_{f}; \ell, m_\ell; M_F}$, where $m_f=m_{f_\text{Rb}}+m_{f_\text{Cs}}$, the
number $\ell$ is the quantum number for the mechanical rotation of the
molecule, and $m_\ell$ is its projection on the laboratory axis. We use the
symbol $\tilde{v}$ to label the vibrational level, numbered downwards from the
Rb($5S_{1/2},f_{Rb}$)+Cs($6S_{1/2}, f_{Cs}$) hyperfine dissociation limit. In
contrast to the homonuclear case, the quantum number $f$
($\vec{f}=\vec{f}_\text{Rb}+\vec{f}_\text{Cs}$) is not a good quantum number.
However, the projection $M_F$ of the total angular momentum
$\vec{F}=\vec{f}+\vec{\ell}$ is always conserved.\cite{Takekoshi2011} The two
initial states $\ket{1\rm{a}}$ and $\ket{1\rm{b}}$ have both $d$-wave
($\ell=2$) character and vibrational quantum number $\tilde{v}\!=\!-6$, which
is the energetically lowest vibrational state that we can currently access with the magneto-association technique.\cite{Takekoshi2011} The full set of quantum
numbers for the two relevant Feshbach levels is $\tilde{v}\!=\!-6;
f_\text{Rb}\!=\!2,m_{f_\text{Rb}}\!=\!2;f_\text{Cs}\!=\!4,m_{f_\text{Cs}}\!=\!2;
m_{f}\!=\!4;\ell\!=\!2,m_\ell\!=\!0; M_F\!=\!4$ for $\ket{1\rm{a}}$ and
$\tilde{v}=-6; f_\text{Rb}=2, m_{f_\text{Rb}}=2; f_\text{Cs}=4,
m_{f_\text{Cs}}=4; m_{f}=6; \ell=2, m_\ell=-2; M_F=4$ for $\ket{1\rm{b}}$. Level $\ket{1\rm{a}}$ effectively has $J=2$ and $M_J=1$, see Section \ref{subsec:Meas0plus}. Level $\ket{1\rm{b}}$ effectively has $J=2$ and $M_J=-1$.

\begin{figure*}
\begin{center}
\includegraphics[width=\textwidth]{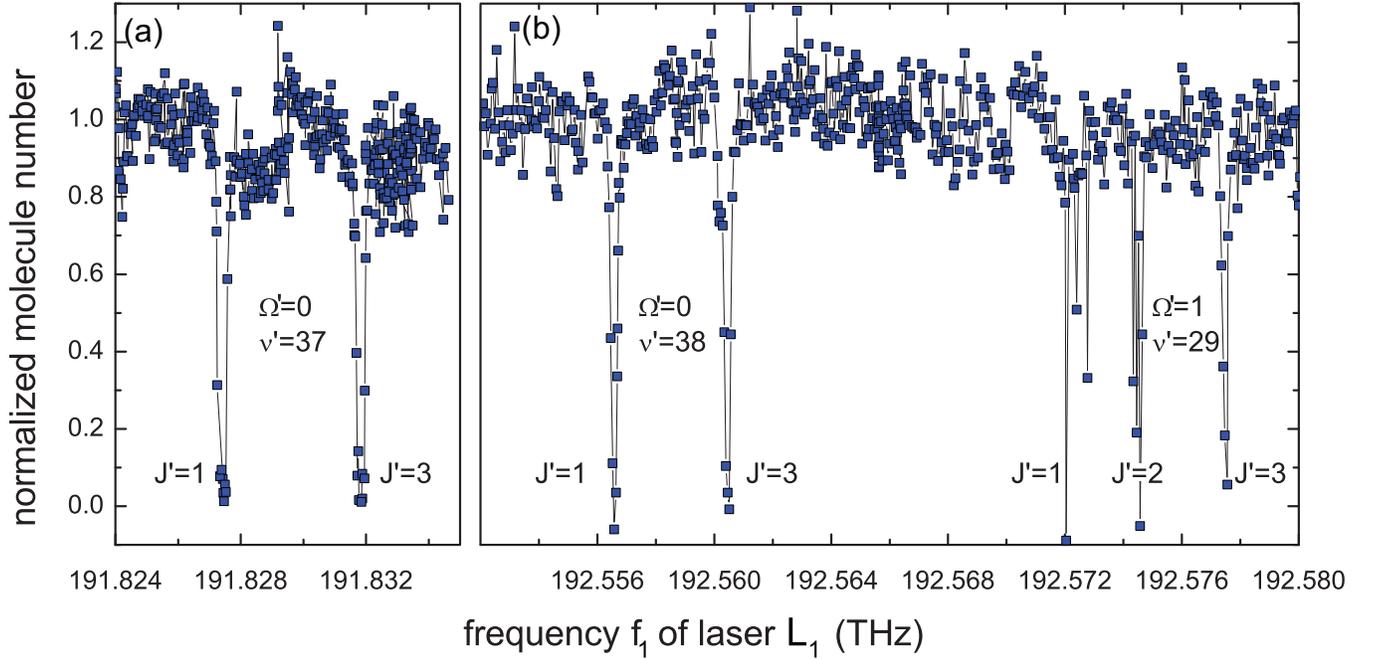}
\caption{Typical loss resonances for excitation from the Feshbach level
$\ket{1\rm{a}}$ to (a) the $v'\!=\!37$ and (b) to the $v'\!=\!38$ levels of the $0^+$ potentials. The Feshbach molecules are irradiated for 10~ms with laser light at an intensity of $I=5\times 10^5$mW/cm$^2$ and we plot the number of remaining Feshbach molecules as a function of the laser frequency $f_1$ of laser $L_1$. For each vibrational level we identify two resonances corresponding to $|J'\!=\!1\!\rangle$ and $|J'\!=\!3\!\rangle$. The half right of panel (b) shows additional loss peaks corresponding to transitions that we attribute to the excited $\Omega'=1$ state. Here we observe a more complex structure that includes rotational and hyperfine levels, as discussed in Sec.\,\ref{subsec:OmegaOne}. The lines are not fits but are guides to the eye and simply connect the data points.} \label{fig:lossres}
\end{center}
\end{figure*}

The light for driving the lambda-type three-level system is derived from two
lasers, $L_1$ and $L_2$. These are widely tunable diode laser systems with
linewidths of about 10 kHz. For short term stability the lasers are locked to
narrow band optical resonators. For long term stability the optical resonators
are referenced to a $^{133}$Cs-vapor saturation spectroscopy signal using
transfer diode lasers and the beat locking technique. We determine the laser
wavelengths by using two independent wavemeters, whose accuracies we estimate
to be about $0.0016$cm$^{-1}$ and $0.0003$cm$^{-1}$. The beams of lasers $L_1$
and $L_2$ have $1/e^2$-intensity radii of 39(1)~$\mu$m and 20(1)~$\mu$m at the
location of the molecular sample, respectively. The laser beams lie in the
horizontal plane and propagate almost collinearly. The magnetic field defining
the quantization axis is oriented along the vertical direction. The light is
polarized either in the horizontal plane ("h", $\epsilon=\pm1$) or in the
vertical ("v", $\epsilon=0$) direction. The light can thus induce either
$\sigma^+/\sigma^-$ or $\pi$ transitions.

\section{The intermediate state: the electronically excited
states of $\Omega'=0^+$ and $\Omega'=1$ symmetry}
\label{sec:0plus}

For ground-state transfer we are primarily interested in transitions from
$\ket{1}$ to rovibronic levels $\ket{v',J'}$ of the electronically excited
mixed states $0^+$. But also transitions to the state with $\Omega'=1$ could be of interest. With our current laser setup we address different vibrational levels ranging from $v'=33$ to $v'=38$ for $0^+$. For each vibrational level we record the excitation spectrum and we measure the single-photon excitation rate.

\subsection{Loss resonances and excitation rates to $\Omega'=0^+$ levels}
\label{subsec:Meas0plus}

In a first set of experiments we perform optical spectroscopy using state
$\ket{1\rm{a}}$. The molecules are irradiated by laser $L_1$ for a certain
irradiation time $t_\text{irr}$, typically ranging from 10~$\mu$s to 10~ms, and we record the number of remaining molecules in $\ket{1\rm{a}}$. For each data point in the spectrum, a new sample is prepared and the laser frequency $f_1$ is stepped to a new value. Since prior to our experiments the precise energies of the excited state levels had poorly been known, we first performed our search over a broad range of frequencies $f_1$ using a comparatively high laser intensity.

\begin{figure*}
\begin{center}
\includegraphics[width=1\textwidth]{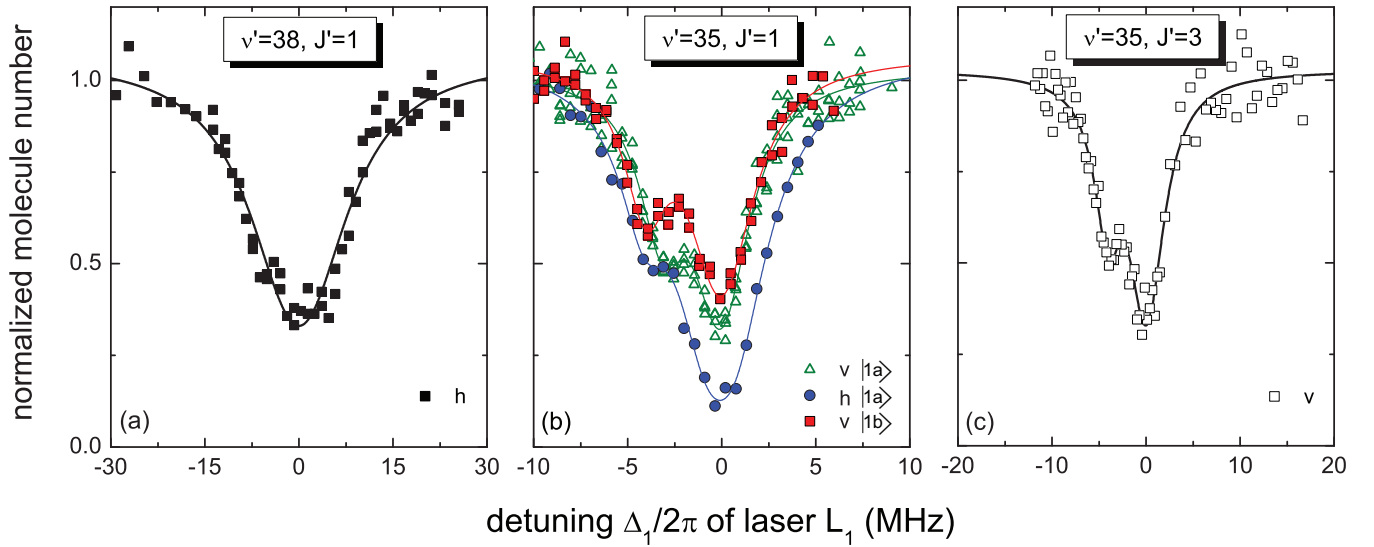}
\caption{High resolution spectroscopy of deeply bound excited levels of RbCs:
Typical loss resonances corresponding to various levels of the $0^+$ states for weak excitation from the initial Feshbach level $\ket{1\rm{a}}$. (a)
$\ket{v'\!=\!38,J'\!=\!1}$, (b) $\ket{v'\!=\!35,J'\!=\!1}$, and (c) to the
$\ket{v'=35,J'=3}$. For comparison, panel (b) also shows a loss feature for
excitation from the level $\ket{1\rm{b}}$ using vertically polarized light (red squares). All measurements are performed with an irradiation time of
$t_\text{irr}=10 \mu$s. The solid lines are fits according to the model discussed in the text. Zero detuning is defined by the minimum of the single exponential fit according to Eq.~\ref{eqn:N}.} \label{fig:peak}
\end{center}
\end{figure*}

We observe a number of resonant loss features caused by excitation to the
electronically excited $0^+$ states. Typical molecular excitation spectra are displayed in Fig.\,\ref{fig:lossres}(a) and (b). The comparison with the calculated energy positions and rotational constants from Ref.\cite{Docenko2010} results in an assignment to the levels $v'=33$ to $v'=38$. As expected, the vibrational levels are not evenly spaced. For each $0^+$ vibrational level $v'$ we detect a two-peak structure, which we ascribe to the $J'\!=\!1$ and $J'\!=\!3$ rotational levels. Fig.\,\ref{fig:peak} presents high resolution scans of selected resonances. Except for the level $v'\!=\!35$, we do not observe any additional substructure, i.e. substructure caused by Zeeman
or hyperfine interaction, in agreement with the expectation of a small
hyperfine splitting of the $0^+$ lines. Table \ref{tab:spectro} summarizes the
transition energies, rotational constants, and further parameters to be
discussed below for the relevant vibrational levels of the $0^+$ potential. We obtain a difference of about $h \times 7.5$~GHz between calculated\cite{Fellows1999} and observed transition energies. This discrepancy results from insufficient knowledge of the $X^1\Sigma^+$ well depth. Since we measure that energy as described below by two-photon spectroscopy (see Section \ref{sec:TwoPhot}), we provide adjusted values for the calculated transition energies.

Only two rotational lines are observed for each of the $0^+$ vibrational transitions shown above. This observation gives us information about $J$ for $\ket{1}$ as well as about the parity selection rules. The Feshbach level $\ket{1}$ has $\ell=2$ and has almost full triplet character.\cite{Takekoshi2011} With $S=1$ and $L=0$ the total angular momentum number can take values $J=1,2,3$. For
level $\ket{2}$, $J'$ is a good quantum number and rotational states
$J'=0,1,2,3,4$ could be populated in principle. However, as no $J'=0,2,4$ lines are observed, we conclude that level $\ket{1\rm{a}}$ has mostly $J=2$ character (the same observation holds for level $\ket{1\rm{b}}$). From the two-photon spectroscopy below we conclude that $M_J=+1$ for level $\ket{1\rm{a}}$. Further, the parity $P_\text{tot}=P_\text{elec}P_\text{rot}$ of the total wave function is a product of the electronic wave function parity ($P_\text{elec}=+1$ for both $0^+$ and $a^3\Sigma^+$ states) and the rotational wave function parity ($P_\text{rot}=(-1)^J$). As is well known, dipole transitions obey the $+1\rightarrow -1$ selection rule for the total parity. Hence, we have $P(\ket{1})=+1$, so that only transitions to level $\ket{2}$
with negative parity are allowed, namely to $J'=1,3$.

We get an estimate for the natural linewidth $\Gamma$ of the excited levels from the high resolution scans shown in Fig.\,\ref{fig:peak}. These are taken with a much smaller frequency step size for higher spectral resolution and a much lower laser intensity to avoid artificial line broadening due to depletion of the initial state. In a first approximation, we extract $\Gamma$ by using a simple two-level model. In the limit of $\Omega_1 \ll \Gamma$ the number of Feshbach molecules $N$ decays according to the following equation
\begin{equation}
\label{eqn:N} N(t_\text{irr}) = N_0 \ \exp{(- t_\text{irr}\Omega_1^2
\Gamma/(\Gamma^2+4\Delta_1^2))},
\end{equation}
where $N_0$ is the initial molecule number, $t_\text{irr}$ as before is the
irradiation time, $\Omega_1$ is the Rabi frequency of $L_1$, and $\Delta_1$ is
the laser detuning. For fixed $t_\text{irr} \Omega_1^2$ we fit
Eq.\,\ref{eqn:N} to the spectroscopic data as a function of $\Delta_1$ and we
extract $\Gamma$. As an example, Fig.\,\ref{fig:peak}(a) shows the loss feature for excitation from $\ket{1\rm{a}}$ to $\ket{v'\!=\!37,J'\!=\!1}$ together with the fit. The excitation spectra for the $\ket{v'\!=\!35,J'\!=\!1}$ and $\ket{v'\!=\!35,J'\!=\!3}$ levels are special cases, exhibiting double loss features as shown in Fig.\,\ref{fig:peak}(b) and (c). We fit these resonances using Eq.\,\ref{eqn:N} generalized to the case of two exponential functions. The presence of two closely spaced loss features is probably the result of hyperfine splitting of $\ket{v'\!=\!35,J'\!=\!1}$ and $\ket{v'\!=\!35,J'\!=\!3}$. It is unlikely that a level of the $\Omega'=1$ state accidentally overlaps with either the $\ket{v'\!=\!35,J'\!=\!1}$ or the $\ket{v'\!=\!35,J'\!=\!3}$ level, as we think that we have identified all the relevant $\Omega'=1$ levels in the wavelength region of interest (see the discussion below in Section \ref{subsec:OmegaOne}). Nevertheless, a further investigation is needed to clarify this issue.

In Table \ref{tab:spectro} we list the measured linewidths that we determine by the single exponential function fits. We note that the values given are only approximate as unresolved hyperfine and Zeeman substructure might lead to apparent line broadening. The computed linewidths are derived from the MFGH wave functions and the $R$-dependent TDMs (Section \ref{sec:molpot}) neglecting hyperfine structure. The singlet-dominated $0^+$ levels, i.\,e.\,the levels $v'\!=\!33,36,$ and $38$, have linewidths of about $\Gamma^\text{calc}\approx 2\pi\times 5$ MHz and the triplet-dominated levels have linewidths of about $\Gamma^\text{calc}\approx 2\pi\times 2$ MHz. The measured linewidths are approximately $1$ MHz larger than the calculated values.

\begin{figure*}
\begin{center}
\includegraphics[width=1\textwidth]{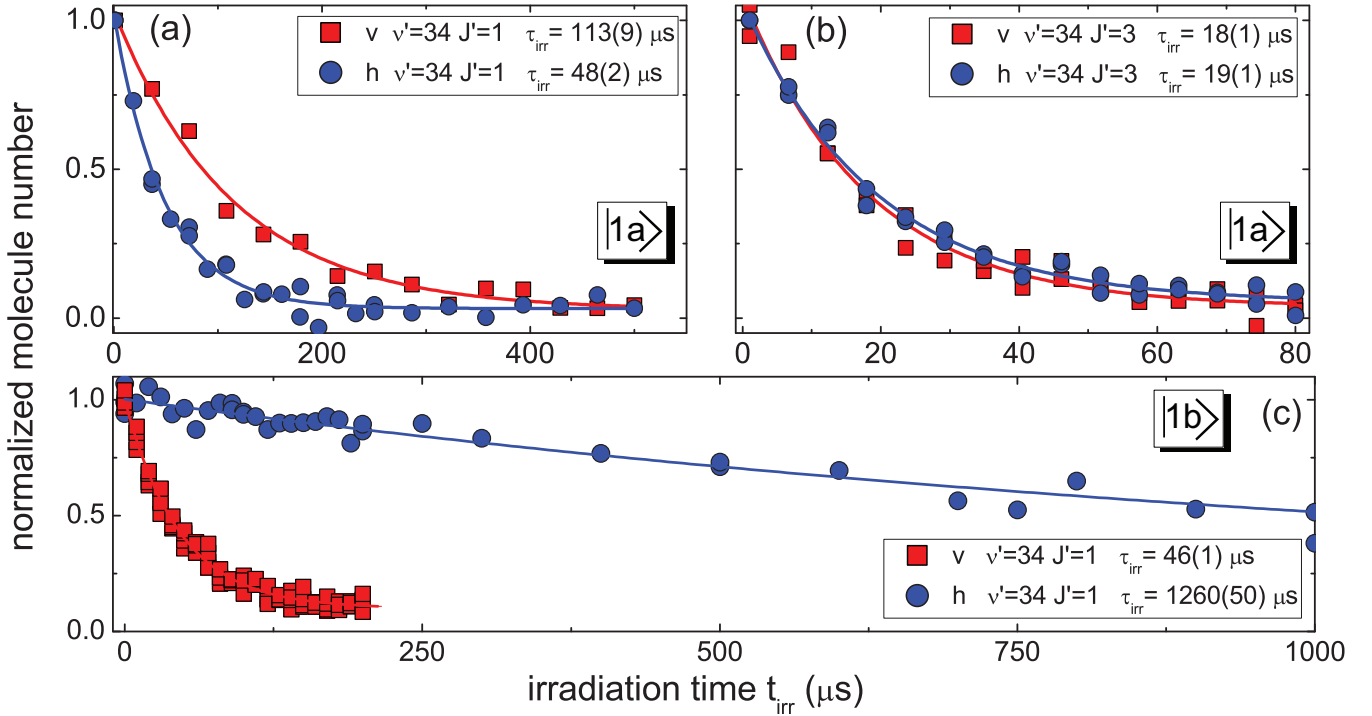}
\caption{Time evolution of the number of Feshbach molecules as a function of
the irradiation time $t_\text{irr} $ for horizontal (h, circles) and vertical
(v, squares) polarization of laser $L_1$. For the initial state we choose level $\ket{1\rm{a}}$ in (a) and (b) and level $\ket{1\rm{b}}$ in (c). The laser
light resonantly couples to $|v'\!=\!34, J\!=\!1\!\rangle$ in (a) and (c) and
to $|v'\!=\!34, J\!=\!3\!\rangle$ in (b) with an intensity of $4.4\times
10^5\rm{mW}/\rm{cm}^2$. We fit Eq.\,\ref{eqn:N} with $\Delta_1=0$ to the data
including a non-zero offset and determine the decay constants $\tau_\text{irr}$ as indicated. The non-zero offset accounts for for residual molecules and for non-perfect background subtraction for the absorption images.}
\label{fig:lifetime}
\end{center}
\end{figure*}

In a second experiment we measure the single photon excitation rates to obtain
an estimate for the Rabi frequency $\Omega_1$ and hence for the transition strength for the various transitions. Again we record the time evolution of the Feshbach-molecule number $N$ as a function of $t_\text{irr}$. This time we keep $L_1$ on resonance. In this case Eq.\,\ref{eqn:N} reduces to $N(t_\text{irr}) = N_0 \ \exp{(- t_\text{irr} \Omega_1^2/\Gamma)}$. Typical decay curves of the molecule number in levels $\ket{1\rm{a}}$ and $\ket{1\rm{b}}$ are shown in Fig.\,\ref{fig:lifetime}. From the fit to the data we determine the decay constant $\tau_\text{irr}=\Gamma/\Omega_1^2$ to obtain the Rabi frequency $\Omega_1=(\Gamma/\tau_\text{irr})^{1/2}$. The linewidth $\Gamma$ is taken from the loss resonance data as described above.

\begin{figure}[h]
\begin{center}
\includegraphics[width=0.45\textwidth]{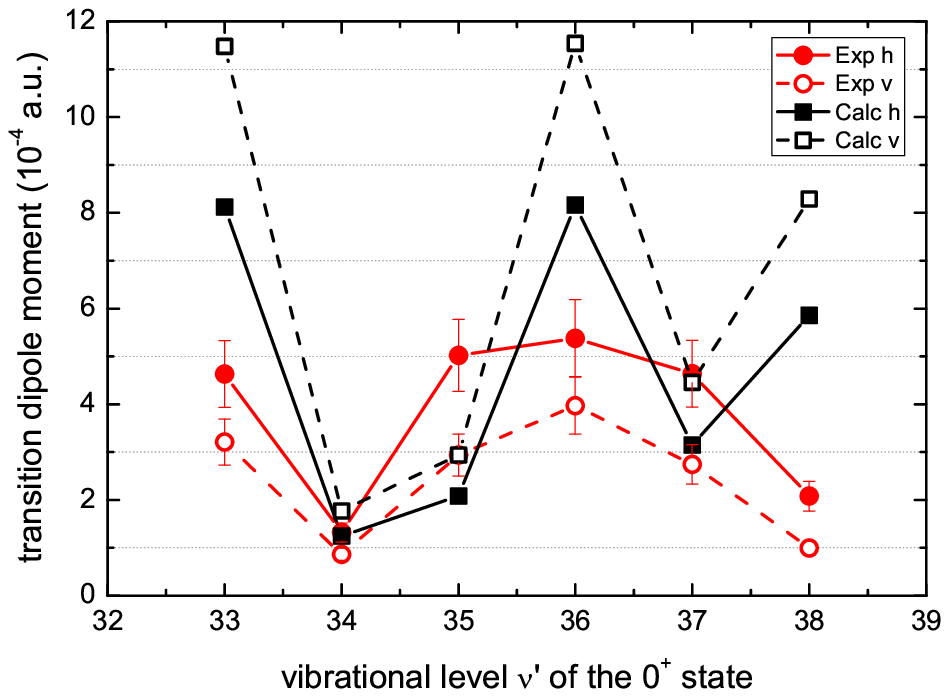}
\caption{Transition dipole moments measured for different vibrational levels
$\ket{v', J'\!=\!1}$ of the $0^+$ state with horizontally polarized (filled
circles) and vertically polarized light (empty circles) in comparison to the
computed values for both horizontally (filled squares) and vertically polarized light (empty squares). The experimental data refer to the initial Feshbach
level $\ket{1\rm{a}}$. In the model, we chose the $v\!=\!-3$ level of the
$a^3\Sigma^+$ potential to match the overall magnitude of the experimental
TDMs, which is equivalent to choosing the $v\!=\!-6$ level scaled by a factor
2.42 (see Fig.\ref{fig:dip_vibration}).} \label{fig:dipmom}
\end{center}
\end{figure}

In Table \ref{tab:spectro} we list the normalized Rabi frequencies determined as described above for the transitions from level $\ket{1\rm{a}}$. We have factored out the dependence on laser intensity $I$. We translate the values into experimental TDMs, which we display in Fig.~\ref{fig:dipmom}. For reference, also computed TDMs are displayed. These come with an arbitrary scaling factor, as discussed in the previous section. For the transitions to $\ket{2} = |v', J'\!=\!3\rangle$ the TDMs are essentially independent of polarization. This agrees with the expectation as the calculated TDMs in this case differ by only a factor $1.1$ in view of the H\"onl-London factors (see Table \ref{tab:Honl-London} for $M_J=+1$ of level $\ket{1\rm{a}}$). However, for the transitions to $\ket{2} = |v', J'\!=\!1\rangle$ we find reduced values for the TDMs when we switch from horizontal to vertical polarization. This is in contrast to the expectation, which gives an increase by a factor $\sqrt{2}$. The discrepancy is even more striking for transitions from the Feshbach level $\ket{1\rm{b}}$ to $\ket{v',J'\!=\!1}$. When we switch from vertical to horizontal polarization the TDM is reduced by more than a factor of $5$. Here, one would again expect an increase by a factor of $\sqrt{2}$. The reduction for the excitation rate by more than a factor of $25$ can clearly be seen in the decay measurements (see Fig.\,\ref{fig:lifetime}(c)). It is most probable that the mismatch between experiment and theory cannot be explained within a two-level model, and we speculate that mixing with additional excited levels could be responsible for the unusual behavior. In future, we plan to investigate this issue in detail as it is important for a full understanding of the molecular structure relevant to the ground-state transfer.

%We also provide the experimental TDMs to allow comparison with the calculated TDMs. In general, we find reasonable agreement between the experimental and the calculated TDMs for transitions starting from $\ket{1\rm{a}}$. We compare these in Fig.~\ref{fig:dipmom}. For the excited state levels with $v'=34, 35$, and $37$ (these have mostly triplet character) they agree within a factor of 2. For the levels $v'$ with mostly singlet character the agreement is within an order of magnitude.

%This pattern cannot be reproduced by our theoretical model (see Figure \ref{fig:dipmom}), which predicts excitation rates that are {\em lower} by a factor of $\sqrt{2}$ for horizontally polarized light than for vertically polarized light, as reflected by the  H\"onl-London factors listed in Table \ref{tab:Honl-London}.

\begin{table*}[ht]
\centering \label{TDM}
\begin{tabular}{|c|c|c|c|c|c|c|c|c|c|c|c|}
\hline $v^{'}$ & $B^\text{calc}_{v'}$
&$B^\text{exp}_{v'}$&$J'$&$E^\text{calc}_{\ket{1\rm{a}},\ket{2}}$ &
$E^\text{exp}_{\ket{1\rm{a}},\ket{2}}$  & $\mathcal{P}$ &
$D^\text{calc}_{\ket{1\rm{a}},\ket{2}}$ &
$D^\text{exp}_{\ket{1\rm{a}},\ket{2}}$  & $\Omega^\text{exp}_{1}$ & $\Gamma^\text{calc}/2\pi$&  $\Gamma^\text{exp}/2\pi$\\
\hline \hline
33               &0.0127&      & 1 &6321.84&6321.850& v & 11.48 & 3.2 & 0.36&5.7&\\
(14.1,85.9)&      &      & &       &        & h & 8.12  & 4.6 & 0.51&5.7&6.5\\
34 &0.0149&0.0150& 1 &6329.36&6329.357& v & 1.77  & 0.9 & 0.10&1.9&\\
(65.2,34.8)&      &      &   &       &        & h & 1.25  & 1.3 & 0.15&1.9&2.9\\
           &      &      & 3 &6329.51&6329.507& v & 3.09  & 2.2 & 0.24&1.9&  \\
           &      &      &   &       &        & h & 2.79  & 2.1 & 0.23&1.9&  \\
35               &0.0147&0.0148& 1 &6364.04&6364.031& v & 2.94  & 2.9 &
0.33&2.2&\\ (59.9,40.1) &      &      &   &       &        & h & 2.08  & 5.0 &
0.56&2.2&4.0\\
            &      &      & 3 &6364.18&6364.179& v & 5.13  & 4.6 & 0.51&2.2&
            \\
            &      &      &   &       &        & h & 4.63  & 4.0 & 0.45&2.2&
            \\
36               &0.0130&      & 1 &6372.30&6372.314& v & 11.54 & 4.0 &
0.44&5.3& \\ (20.3,79.7) &      &      &   &       &        & h & 8.16  & 5.4
& 0.60&5.3&6.5\\ 37          &0.0147&0.0148& 1 &6398.67&6398.663& v & 4.46  &2.7
& 0.30  &2.0&\\ (60.0,40.0) &      &      &   &       &        & h & 3.15  &
4.6 & 0.52&2.0&3.1\\
            &      &      & 3 &6398.82&6398.811& v & 7.78  &    &      &2.0&
            \\
            &      &      &   &       &        & h & 7.02  &    &      &2.0&
            \\
38          &0.0130&      & 1 &6422.97&6422.986& v & 8.29  & 1.0& 0.11&5.3&\\
(21.3,78.7) &      &      &   &       &        & h & 5.86  & 2.1 & 0.23&5.3&5.8\\
\hline
\end{tabular}
\caption{\label{tab:spectro} Vibrational quantum numbers $v'$ (with the
percentage of $b^3\Pi$ and $A^1\Sigma^+$ character in parentheses), experimental and calculated rotational constants $B_{v'}$ (in units cm$^{-1}$), experimental and calculated transition energies $E_{\ket{1\rm{a}},\ket{2}}$ (in units cm$^{-1}$), experimental and calculated transition dipole
moments (TDMs) $D_{\ket{1\rm{a}},\ket{2}}$ (in units $10^{-4}$ a.u.), normalized Rabi frequencies $\Omega^\text{exp}_{1}$ (in units kHz $\times (I\rm{/(mW/cm^2)})^{1/2}$, where $I$ is the laser intensity), and experimental and calculated natural linewidths $\Gamma/(2\pi)$ (in units MHz) for the transitions from level $\ket{1\rm{a}}$ to the $0^+$ levels in the wavelength range from about 1581~nm to 1557~nm for $J'=1,3$ and for vertical (v, $\epsilon=0$) and horizontal (h, $\epsilon=\pm 1$) polarization $\mathcal{P}$. The calculated transition energies are extracted from Ref.\cite{Docenko2010} and corrected using our new determination of the $X^1\Sigma^+$ ground-state energy, see Section \ref{sec:TwoPhot} below. The experimental transition energies are corrected for the Zeeman shift by extrapolation to zero magnetic field. The $1\sigma$-statistical errors are $0.0002$cm$^{-1}$ for $B^\text{exp}_{v'}$, $0.002$cm$^{-1}$ for $E^\text{exp}_{\ket{1\rm{a}},\ket{2}}$, and 15\% for
$D^\text{exp}_{\ket{1\rm{a}},\ket{2}}$, $\Omega^\text{exp}_{1}$,
and $\Gamma^\text{exp}$.}
\end{table*}

\subsection{Single photon spectroscopy of the $\Omega'=1$ excited state}
\label{subsec:OmegaOne}
In principle, excitation to levels of the electronically excited $\Omega'=1$ and $\Omega'=2$ potentials is possible starting from our Feshbach level. Some levels of the electronically excited $\Omega'=0$ and $\Omega'=1$ potentials are quite close to each other and could even accidentally overlap. It is thus very important for the future STIRAP transfer scheme to be able to distinguish between these two classes of levels. In the course of our spectroscopic measurements we have detected a number of levels belonging to the $\Omega'=1$ potential. In Fig.~\ref{fig:lossres} (b) we clearly see additional loss features on the high-frequency side of the $v'=38$ resonances. These peaks cannot be attributed to any level of either the $\Omega'=0$ or the $\Omega'=2$ potential and hence must be caused by excitations to levels of the $\Omega'=1$ potential.\cite{Docenko2010} Based on the energy and rotational splittings we identify this level to be the vibrational level with $v'=29$. The loss peaks exhibit a much richer and a more complex substructure than the $\Omega'=0$ features (the substructure is not clearly resolved in the coarse scan shown in Fig.~\ref{fig:lossres} (b)). This behavior is to be expected as a result of the excited state Zeeman and hyperfine splitting. Note that the rotational structure is a multiplet of $J'=1, 2, 3$ lines. Excitation to $J'=2$ is allowed by the parity selection rules discussed above, because the $\Omega'=1$ potential is doubly degenerate with two individual components of $+1$ and $-1$ parity.

\label{sec:TwoPhot}
\begin{figure*}
\begin{center}
\includegraphics[width=1\textwidth]{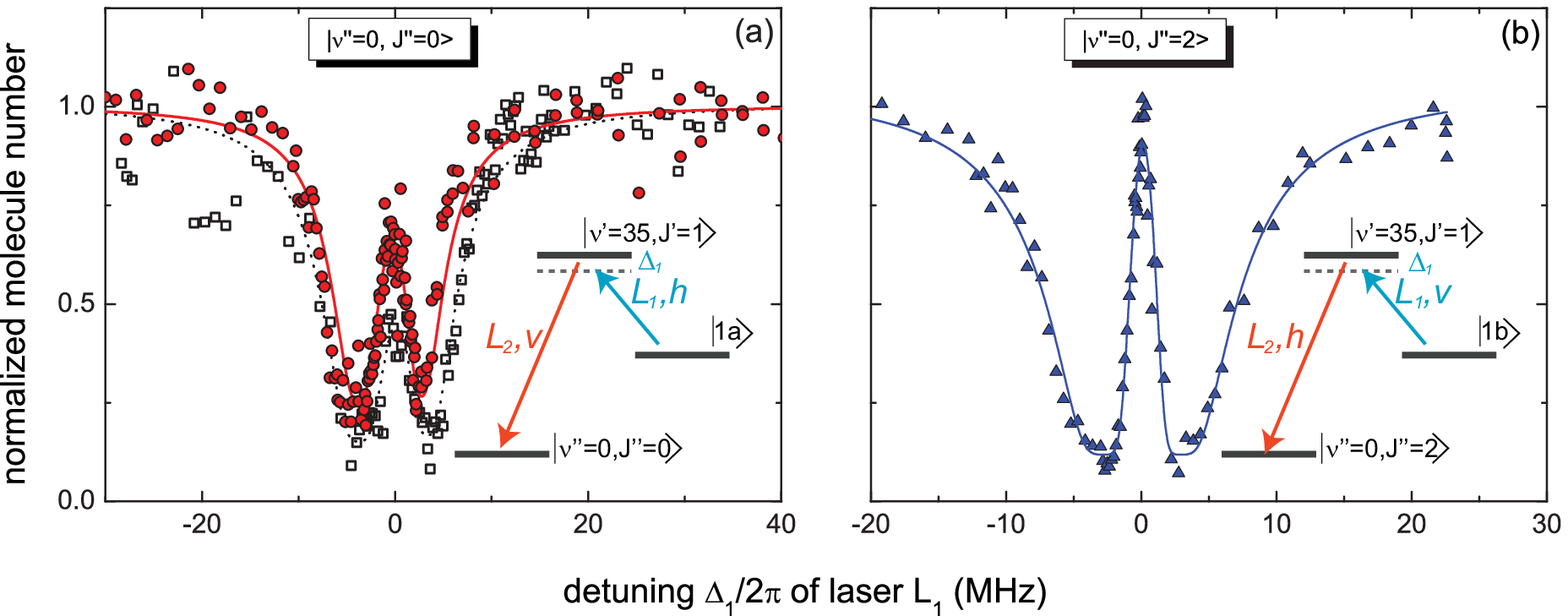}
\caption{Dark resonances involving the ground-state vibrational level
$\ket{v''\!=\!0}$ of the $X^1\Sigma^+$ state. We plot the relative number of
remaining Feshbach molecules as a function of the detuning $\Delta_1$ of laser $L_1$. Laser $L_2$ is on resonance. The strong suppression of loss at zero detuning reflects the creation of the dark state. The solid lines are fits to the data using a three-level model that includes loss processes. For details see text. (a) The dark resonances involve the initial Feshbach level $\ket{1\rm{a}}$, the intermediate level $\ket{2}=\ket{v'\!=\!35, J'\!=\!1}$, and the final level $\ket{3}\ket{v''\!=\!0,J''\!=\!0}$. The polarizations of lasers $L_1$ and $L_2$ are chosen to be h and v, respectively. Lasers $L_1$ and $L_2$ simultaneously irradiate the sample for 20~$\mu$s (circles) and 40~$\mu$s (squares). (b) Here, the initial Feshbach level is $\ket{1\rm{b}}$, the intermediate level is $\ket{2}=\ket{v'\!=\!35, J'\!=\!1}$, and the final level is $\ket{v''\!=\!0,J''\!=\!2}$. Lasers $L_1$ and $L_2$ have v and h polarization and are directed onto the molecular sample for 100~$\mu$s. Zero detuning is defined by the center of the one-photon resonance derived from a single exponential fit according to Eq.\ref{eqn:N}.} \label{fig:EIT}
\end{center}
\end{figure*}

\section{Two-photon dark state resonance spectroscopy}
\label{sec:TwoPhot}
The next step towards coherent optical transfer into the rovibronic ground state is the search for a suitable dark state resonance. It is well known that an essential property of a lambda-type three-level system is the existence of a dark state as an eigenstate of the system.\cite{Bergmann1998cpt} This state is dark in the sense that it is decoupled from the excited state $\ket{2}$. It is thus not directly influenced by the excited state's radiative decay. The existence of such a loss-free state lies at the heart of the STIRAP technique. It is thus of prime importance to find an appropriate dark state resonance. Here we consider the three-level system involving one of the initial Feshbach states, i.e. either $\ket{1\rm{a}}$ or $\ket{1\rm{b}}$, an intermediate level $\ket{2}\!=\!\ket{v',J'\!=\!1}$ belonging to the $0^+$ coupled potentials, and the rovibronic ground-state level $\ket{3}=\ket{v''\!=\!0,J''\!=\!0}$ as the final level. Alternatively, we take $\ket{3}=\ket{v''\!=\!0,J''\!=\!2}$ as the third level and compare the data to the data involving the rovibronic ground-state level.

In the experiment we prepare the Feshbach molecules either in state $\ket{1\rm{a}}$ or in state $\ket{1\rm{b}}$ and simultaneously irradiate the sample with rectangular light pulses from $L_1$ and $L_2$. Typical irradiation times range between 10~$\mu$s to 100~$\mu$s. The laser light intensities are chosen such that the Rabi frequency $\Omega_2$ for $L_2$ is much greater than $\Omega_1$, the Rabi frequency for $L_1$. Immediately after irradiation we probe the molecule number. We repeat this procedure as we vary the detuning $\Delta_1$ of laser $L_1$, whereas $L_2$ is kept in resonance ($\Delta_2=0$). In this way we have explored all the dark state resonances that involve the various excited levels $\ket{2}\!=\!\ket{v',J'\!=\!1}$ found in the one-photon spectroscopy measurements discussed above. We find that the excited level $\ket{2}\!=\!\ket{v'\!=\!35,J'\!=\!1}$ so far gives us the best results as it has the highest two-photon transition strength. Typical dark state spectra for coupling to either the $\ket{v''\!=\!0,J''\!=\!0}$ level or to the $\ket{v''\!=\!0,J''\!=\!2}$ level are shown in Fig.\,\ref{fig:EIT}(a) and (b). The strong suppression of loss at zero detuning is caused by the appearance of the dark state. In the limit of $\Omega_1 \ll \Omega_2$ there is an analytic solution for the lambda-type three-level system\cite{Fleischhauer2005EIT} that allows us to perform a simple fit of the w-shaped spectrum in the form of
\begin{equation}
\label{eqn:model}
N=N_0 \ \exp{\left(-t_\text{irr}\Omega_1^2\frac{(4\Gamma \Delta^2+\Gamma_\text{eff}
(\Omega_2^2+\Gamma_\text{eff}\Gamma)}{|\Omega_2^2+(\Gamma+2i\Delta_1)(\Gamma_\text{eff}+2i\Delta)|^2}\right)}.
\end{equation}
Here, $\Delta=\Delta_1-\Delta_2$ is the two-photon detuning and the parameter $\Gamma_\text{eff}$ phenomenologically accounts for loss and  decoherence between levels $\ket{1}$ and $\ket{3}$. From the fit we find for the transition from $\ket{1\rm{a}}$ to  $\ket{v''\!=\!0,J''\!=\!0}$ that $\Omega_1=2\pi\!\times\! 0.4(3)$MHz and $\Omega_2=2\pi\!\times\!6.3(6)$MHz, resulting in $\Omega_1\!=\!2\pi\!\times\!0.6(4)$kHz$(I/(\text{mW/cm}^2))^{1/2}$ and $\Omega_2\!=\!2\pi\!\times\!7.4(7)$kHz$(I/(\text{mW/cm}^2))^{1/2}$. For the transition from $\ket{1\rm{b}}$ to $\ket{v''\!=\!0,J''\!=\!2}$ we find that $\Omega_1=2\pi\!\times\! 0.16(2)$MHz and $\Omega_2=2\pi\!\times\!5.7(3)$MHz, resulting in $\Omega_1\!=\!2\pi\!\times\!0.54(8)$kHz$(I/(\text{mW/cm}^2))^{1/2}$ and
$\Omega_2\!=\!2\pi\!\times\!6.7(3)$kHz$(I/(\text{mW/cm}^2))^{1/2}$. These values,
given sufficient laser coherence, should be good enough for performing STIRAP
transfer to the vibrational ground state. We note that we measure a
comparatively short lifetime for the dark state when we couple to the
$\ket{v''\!=\!0,J''\!=\!0}$ level. Fig.\,\ref{fig:EIT}(a) shows that the dark
state involving $\ket{v''\!=\!0,J''\!=\!0}$ decays on the timescale of $10
\mu$s. A non-zero $\Gamma_\text{eff}$ is responsible for this behavior. In principle, the dark state would show decay if there were not enough
laser coherence. However, our data involving the ground-state level
$\ket{v''\!=\!0,J''\!=\!2}$ (see Fig.\,\ref{fig:EIT}(b)) shows essentially no
decay on the timescale of $100 \mu$s. This fact thus excludes the possibility
of insufficient laser coherence. In principle, the short lifetime could limit
the efficiency of the STIRAP transfer to $\ket{v''\!=\!0,J''\!=\!0}$ level. We
expect that we have to go beyond the simple three-level approximation to
understand the reduced lifetime. Additional investigations are needed to shine
light on this crucial issue.

From the dark state data we determine the binding energy of the lowest rovibronic level $\ket{v''\!=\!0,J''\!=\!0}$ of the ground state as follows. The difference of laser frequencies at 217 G (laser $L_1$ at 6364.0439cm$^{-1}$ and laser $L_2$ at 10175.2913 cm$^{-1}$ to populate the $\ket{v''\!=\!0,J''\!=\!0}$ level) measured with respect to the Rb($5^2S_{1/2}$,$f_{Rb}=1$)+Cs($6^2S_{1/2}$,$f_{Cs}=3$) asymptote yields 3811.2477 cm$^{-1}$. The binding energy at zero field, with respect to the center of gravity of the hyperfine manifold, 3811.5755(16) cm$^{-1}$, is obtained by adding the Zeeman shift (0.0131cm$^{-1}$, or 0.394 GHz) and the relevant energy spacing (0.3150cm$^{-1}$, or 9.443 GHz). The error estimates our wavemeter precision. This result represents a 300-fold improvement in accuracy with respect to the previous value of 3811.3(5) cm$^{-1}$ from Ref.\cite{Fellows1999}. Note that in Table 2, the computed $0^+$ level energies have been corrected by 0.24cm$^{-1}$, as the extraction of potential curves from Ref.\cite{Docenko2010} relies on the latter value. Our measurement is entirely limited by the precision of our wavemeters and could be greatly improved by e.g. referencing the lasers to a frequency comb. The new value will allow more accurate coupled-channel calculations to determine the complete ground-state structure of the RbCs system.

\section{Conclusions and outlook}
We have performed high resolution one- and two-photon laser spectroscopy on
ultracold samples of RbCs molecules with the aim to identify a suitable
two-photon transition for stimulated ground-state transfer. In particular, we
have identified the rovibronic ground state $\ket{v''\!=\!0,J''\!=\!0}$ by
creating a dark state involving the initial Feshbach level and
$\ket{v''\!=\!0,J''\!=\!0}$. With sufficient laser coherence we are now in the
position to perform STIRAP and to create high density samples of ultracold RbCs  ground-state molecules. Presently, we are implementing a 3D optical lattice with the aim to aid the molecule creation and transfer processes by creating a
two-atom Rb-Cs Mott insulator many-body state.

\section*{Acknowledgements}
We thank C. Amiot, T. Bergeman, J. Hutson, P. Julienne, R. Le Sueur, M. J. Mark, and J. G. Danzl for helpful discussions and E. Tiemann for providing a refined $X^1\Sigma^+$ ground-state potential. We acknowledge support by the
Austrian Science Fund (FWF) through the Spezialforschungsbereich (SFB) FoQuS within project P06 (FWF project number F4006-N16). N.B., O.D. and R.V. are members of the \textit{Institut de Recherches sur les Atomes Froids} (IFRAF).

\footnotesize{
\bibliography{RbCs-Spectroscopy} %your .bib file
\bibliographystyle{rsc} %the RSC's .bst file
} % rsc
\end{document}